\documentclass[aps,amsmath,amssymb,nofootinbib,reprint,preprintnumbers]{revtex4-1}
\usepackage{graphicx}%
\usepackage{dcolumn}%
\usepackage{bm}%
\usepackage{hyperref}

\newcommand{\beqa}{\begin{eqnarray}}
\newcommand{\eeqa}{\end{eqnarray}}

\newcommand{\bea}{\begin{eqnarray}}
\newcommand{\eea}{\end{eqnarray}}

\newcommand{\CN}{{\mathcal N}}

\newcommand\SU{{\rm SU}}
\newcommand\U{{\rm U}}

\newcommand\tr{\mathrm{tr}}

\newcommand\diag{\mathrm{diag}}

\def\tr{\mathop{\mathrm{tr}}\nolimits}

\def\SU{\mathrm{SU}}

\def\vev#1{\langle#1\rangle}

\def\SU{\mathrm{SU}}

\def\diag{\mathop{\mathrm{diag}}}
\def\tr{\mathop{\mathrm{tr}}\nolimits}
\def\vev#1{\langle#1\rangle}

\def\diag{\mathop{\rm diag}\nolimits}

\def\SU{\mathop{\rm SU}}
\let\U\UU

\def\tr{\mathop{\rm tr}}

\def\beq#1\eeq{\begin{align}#1\end{align}}

\newcommand{\beqr}{\begin{array}}
\newcommand {\eeqr}{\end{array}}

\begin{document}

\title{Dynamical Supersymmetry Breaking with $T_{N}$ Theory
}

\preprint{CALT-68-2852, IPMU-13-0145, UT-13-28}

\author{Kazunobu Maruyoshi}
\affiliation{California Institute of Technology,  Pasadena, CA 91125, USA}
\email{maruyosh@caltech.edu}

\author{Yuji Tachikawa}
\affiliation{Department of Physics, Faculty of Science, \\
 University of Tokyo,  Bunkyo-ku, Tokyo 133-0022, Japan,\\
and Institute for the Physics and Mathematics of the Universe (WPI), \\
 University of Tokyo,  Kashiwa, Chiba 277-8583, Japan}
\email{yuji.tachikawa@ipmu.jp}

\author{Wenbin Yan}
\affiliation{California Institute of\ Technology,  Pasadena, CA 91125, USA}
\email{wbyan@theory.caltech.edu}

\author{Kazuya Yonekura}
\affiliation{School of Natural Sciences, Institute for Advanced Study,\\
1 Einstein Drive, Princeton, NJ 08540 USA }
\email{yonekura@ias.edu}

\begin{abstract}
  We demonstrate that the supersymmetry is dynamically broken in the four-dimensional $\SU(N)$ gauge theory coupled to a strongly-coupled superconformal theory $T_N$.
  This is a direct generalization of the model of supersymmetry breaking on deformed moduli space in supersymmetric QCD with an $\SU(2)$ gauge group.
  % This is a direct generalization of the model of the supersymmetry breaking proposed by Izawa and Yanagida~\cite{Izawa:1996pk} and Intriligator and Thomas~\cite{Intriligator:1996pu} which was for $\SU(2)$.
\end{abstract}


\setcounter{tocdepth}{2}
\maketitle
%%%%%%%%%%%%%%%% section 1 %%%%%%%%%%%%%%%%%%%%%%%%%%%%%%%%%%%%%%%%

\section{Introduction}
\label{sec:intro}

Before the discovery of asymptotic freedom in Yang-Mills theory,
there was no ultraviolet (UV) Lagrangian which describes the dynamics of hadrons such as pions and nucleons.
However, the non-existence of Lagrangian did not stop people from studying hadron dynamics.
The assumption of approximate global symmetry $\SU(2)_L \times \SU(2)_R \times \U(1)$ spontaneously broken to $\SU(2) \times \U(1)$
allows us to deduce many properties of hadron dynamics via techniques such as current algebra and effective field theory.
It is also possible to gauge a subgroup of the global symmetry, $\U(1)_{\rm EM} \subset \SU(2)_L \times \SU(2)_R \times \U(1)$, of
such a theory without a Lagrangian (at least in those days).
This construction leads to very important dynamical consequences, such as pion decay to photons, generation of mass difference between charged and neutral pions, etc.
Much of the calculation can be done without the UV  Lagrangian which is QCD.

Similarly, there are theories which have no known Lagrangian description (at least for now) but nevertheless have been found to exist theoretically.
One of them is the $T_{N}$ theory \cite{Gaiotto:2009we} which was discovered in the study of dualities of $\CN{=}2$ supersymmetric theories.
One of the most important properties is that it has the global flavor symmetry $\SU(N) \times \SU(N) \times \SU(N)$ in addition to
some $R$-symmetries. Then we can consider
$\CN{=}1$ supersymmetric systems in which
a subgroup of the global symmetry is coupled to $\CN{=}1$ vector multiplets.
Many  dynamical properties of such theories have been understood.
Instead of experimental data which was available in the case of hadron physics,
we can rely on the power of supersymmetry (SUSY) and duality to study these systems.

In this paper, we demonstrate that these theories without known Lagrangian descriptions are not only theoretically interesting,
but also can have applications in phenomenology. We construct a new model of dynamical supersymmetry breaking using the $T_N$ theory.
In fact, when $N=2$, the $T_{N=2}$ theory is just a set of eight free chiral multiplets in the trifundamental representation of
$\SU(2) \times \SU(2) \times \SU(2) \subset \U(8)$, and
our model is the same as the model of supersymmetry breaking on deformed moduli space of ${\cal N}=1$ supersymmetric QCD
proposed in \cite{Izawa:1996pk,Intriligator:1996pu}.
%Izawa and Yanagida~\cite{Izawa:1996pk} and Intriligator and Thomas~\cite{Intriligator:1996pu}.%\footnote{The authors are agnostic about which of \cite{Izawa:1996pk}  and \cite{Intriligator:1996pu} should be listed first from a proper scientific point of view. They decided to use this order just because of the East-Asian, Confucianist point of view that one of the authors is a direct student of one of the four persons involved in \cite{Izawa:1996pk,Intriligator:1996pu}. }
Therefore our model can be regarded as a direct extension of their model from $\SU(2)$ to $\SU(N)$.
Having such an application to SUSY breaking, it might not be totally nonsense to hope that the $T_N$ theory might be discovered even in
future experiments.

%%%%%%%%%%%%%%%%%%%%%%%%%%%%%%%%%%%%%%%%%%%%%%%%%%%%%%%%%%%%%%%%%%%%%%
\section{The model of supersymmetry breaking}
\label{sec:higgs}
  Let us first recall some properties of the $T_{N}$ theory.
  The $T_N$ theory is an $\CN{=}2$ superconformal field theory (SCFT) which has the global symmetry
  $\SU(N)_A \times \SU(N)_B \times \SU(N)_C \times \U(1)_{R} \times \SU(2)_R$,
  where the last two are the $R$-symmetry of the $\CN{=}2$ superconformal algebra.
  In the language of $\CN{=}1$ supersymmetry, this theory includes, among others, the chiral operators $\mu_A$, $\mu_B$ and $\mu_C$
  transforming in the adjoint representations of $\SU(N)_A$, $\SU(N)_B$ and $\SU(N)_C$, respectively.
They have the scaling dimension two, and satisfy
   the chiral ring relation
    \bea
    \tr \mu_{A}^{k}
     =   \tr  \mu_{B}^{k}
     =     \tr \mu_{C}^{k},
           \label{classicalchiralring}
    \eea
  where $k=2,\ldots, N$.

When $N=2$,  the $T_N$ theory is a set of eight free chiral multiplets $Q_{i_A i_B i_C}$ in the trifundamental
  representation of $\SU(2)_A \times \SU(2)_B \times \SU(2)_C$, and  the  operators $\mu_{A}$ are given by
  $(\mu_A)_{i_A}^{j_A}=Q_{i_A i_B i_C} Q^{j_A i_B i_C}$
  where   the indices of $Q_{i_A i_B i_C}$ are raised and lowered by using the completely anti-symmetric tensor $\epsilon^{i_A j_A}$ etc.
The operators $\mu_{B,C}$ are given in a similar manner, and a direct computation shows that
  the relation \eqref{classicalchiralring} is satisfied in this case.
  However, for $N \geq 3$, we cannot interpret $\mu_{A,B,C}$ as being composites of more fundamental operators.

  We consider the $T_{N}$ theory coupled to $\CN{=}1$ vector multiplet
  by gauging the $\SU(N)_{C}$ flavor symmetry, and also coupled to two gauge singlet chiral multiplets $M_{X}$ ($X = A,B$)
  transforming in the adjoint representations of $\SU(N)_{X}$ by the following superpotential term
    \bea
    W
     =   - \lambda_{A} \tr M_{A} \mu_{A} + \lambda_{B} \tr M_{B} \mu_{B}. \label{eq:treesuper}
    \eea
  The global non-anomalous symmetry of this UV theory is $\SU(N)_{A} \times \SU(N)_{B} \times \U(1)_{R}$
  where $\U(1)_R$ is the same as the $\U(1)_{R}$ symmetry of the $\CN{=}2$ $R$-symmetry of the $T_{N}$ theory.
  The operators $\mu_{A,B}$ are neutral under this $\U(1)_R$, and
  due to the superpotential, $M_{A,B}$ has charge $2$ under this $\U(1)_{R}$.
  The couplings $\lambda_A$ and $\lambda_B$ are dimensionless
  because the scaling dimensions of $\mu_{A,B}$ are two.
When $N=2$, this superpotential is exactly the one used in \cite{Izawa:1996pk, Intriligator:1996pu}.

  As shown in \cite{Maruyoshi:2013hja}, the operators of the $T_N$ theory  neutral under $\SU(N)_C$
  are polynomials of $\mu_{A}$ and $\mu_{B}$.
  The constraints \eqref{classicalchiralring} are deformed by the dynamics
  of the $\SU(N)_C$ gauge group as
    \bea
    \tr \mu_{A}^{k} - \tr \mu_{B}^{k}
     =     N \Lambda^{2N} \delta^{k,N}. \label{eq:deformedmoduli}
    \eea
  This quantum deformation in the theory without the gauge singlet fields $M_{A,B}$ was proposed
  in \cite{Maruyoshi:2013hja} and checked by 't Hooft anomaly matching and by the consistency
  with the results of $\CN{=}1$ SUSY QCD with $N_{f} = N$ flavors when an appropriate vacuum expectation value is given to $\mu_A$.
  In the case of $N=2$ where $(\mu_A)_{i_A}^{j_A}=Q_{i_A i_B i_C} Q^{j_A i_B i_C}$ and $(\mu_B)_{i_B}^{j_B}=Q_{i_A i_B i_C} Q^{i_A j_B i_C}$,
  one can check that \eqref{eq:deformedmoduli} is just a rewriting of the more familiar deformed moduli constraint   $\det M -B \tilde{B}=\Lambda^4$ of $N_f=N=2$ SUSY QCD~\cite{Seiberg:1994bz}.

  The form of the constraints is not changed by the coupling to the gauge singlet fields
  as can be seen from the $R$-charge assignment.
  Therefore the effective low energy superpotential is
  %\begin{widetext}
    \bea
    W
     &=&   - \lambda_{A} \tr M_{A} \mu_{A} + \lambda_{B} \tr M_{B} \mu_{B} \nonumber \\
    &&\qquad     + \sum_{k=2}^{N} \frac{X_{k}}{k} (\tr \mu_{A}^{k} - \tr \mu_{B}^{k} - N \Lambda^{2N} \delta^{k,N}),
           \label{effective}
    \eea
  %\end{widetext}
  where we have introduced the Lagrange multipliers $X_{k}$.
  The $F$-term conditions of $M_A$ and $M_B$, which require $\mu_A=\mu_B=0$, cannot be satisfied due to the quantum deformation term.
  Thus the supersymmetry is spontaneously broken.
When $N=2$,
  this effective superpotential is exactly the one considered in \cite{Izawa:1996pk, Intriligator:1996pu} and our supersymmetry breaking mechanism is a direct generalization of the one in \cite{Izawa:1996pk, Intriligator:1996pu} to $\SU(N)$.

%%%%%%%%%%%%%%%%%%%%%%%%%%%%%%%%%%%%%%%%%%%%%%%%%%%%%%%%%%%%%%%
\section{Absence of the Run-away}
  To show that the supersymmetry breaking vacuum found in the previous section is stable,
  we need to check that there is no runaway behavior.
\subsection{Pseudo-moduli space}
  First of all, we identify the pseudo-moduli direction by neglecting quantum corrections to the K\"ahler potential of $M_{A,B}$.
  By the contribution of the $F$-components of $M_A$ and $M_B$ in the effective superpotential \eqref{effective},
  the potential is bounded as
    \bea
    &V&
    \geq
    V_{M}
    =   |\lambda_A|^2\tr (\mu_A^\dagger \mu_A) + |\lambda_B|^2 \tr (\mu_B^\dagger \mu_B)
           \nonumber \\
    & &
         + \bigl(\sum_{k=2}^N \frac{F_{k}}{k} \left( \tr \mu_A^k -\tr \mu_B^k-N\Lambda^{2N} \delta^{k,N}
           \right)+{\rm c.c.} \bigr),
           \label{eq:Vinequality}
    \eea
  where $F_k$ is the $F$-component of $X_k$.
  Let us study the minimum of $V_M$,
   obtained by solving
\bea
0%=\frac{\partial V_M}{\partial \mu_{A}}
&=&|\lambda_{A}|^2   \mu_{A}^\dagger+\sum_{k=2}^N F_{k} \left(\mu_{A}^{k-1} -  \frac{{\bf 1}}{N}\tr\mu_{A}^{k-1}  \right),
\label{eq:VMeq1} \\
0&=&
%\frac{\partial V_M}{\partial \mu_{B}}=
|\lambda_{B}|^2   \mu_{B}^\dagger-\sum_{k=2}^N F_{k} \left(\mu_{B}^{k-1} - \frac{{\bf 1}}{N}\tr\mu_{B}^{k-1}   \right),
\label{eq:VMeq2} \\
0&=& %k \frac{\partial V_M}{\partial F_k} =
 \tr \mu_A^k -\tr \mu_B^k-N\Lambda^{2N} \delta^{k,N},
\label{eq:VMeq3}
\eea
and their complex conjugates.
From the above equations, it follows that $[\mu_A^\dagger, \mu_A]=[\mu_B^\dagger, \mu_B]=0$.
  Therefore, $\mu_{A,B}$ can be diagonalized by using the $\SU(N)_{A,B}$ transformations, so that we can parametrize the diagonal components as $\mu_A = \diag (\mu_{A,1},\cdots, \mu_{A,N})$ and
  $\mu_B = \diag (\mu_{B,1},\cdots, \mu_{B,N})$.
  Then, $V_{M}$ is now
%  \begin{widetext}
\bea
%    \begin{equation}
    V_{M}
     &=&     |\lambda_A|^2 \sum_{i=1}^{N} |\mu_{A,i}|^2+|\lambda_B|^2 \sum_{i=1}^{N} |\mu_{B,i}|^2 \nonumber \\
&\geq&  %N|\lambda_A|^2 \left[\prod_{i=1}^{N} |\mu_{A,i}|^2\right]^{\frac{1}{N}}+N|\lambda_B|^2 \left[\prod_{i=1}^{N} |\mu_{B,i}|^2\right]^{\frac{1}{N}} \nonumber \\
N\left(|\lambda_A|^2 |\det \mu_A|^{\frac{2}{N}}+|\lambda_B|^2 |\det \mu_B|^{\frac{2}{N}} \right), \label{eq:lowerbound}
%\end{equation}
\eea
%\end{widetext}
where we have used the general inequality $\frac{1}{N} \sum_{i=1}^N a_i \geq \left(\prod_{i=1}^N a_i \right)^{\frac1N}$,
for $a_i \geq 0$, whose equality holds if and only if $a_1=\cdots=a_N$.

Without loss of generality, we can assume that $|\lambda_A| \leq |\lambda_B|$. Then
one can see that the minimum of the right-hand-side of \eqref{eq:lowerbound} under the constraint
$\det \mu_A -\det \mu_B=(-1)^{N-1}\Lambda^{2N}$, which can be derived from $\tr \mu_A^k -\tr \mu_B^k=N\Lambda^{2N} \delta^{k,N}$,
is at $\det \mu_A=(-1)^{N-1}\Lambda^{2N}$ and $\det \mu_B=0$.
When $|\lambda_A|=|\lambda_B|$, the vacuum with $A \leftrightarrow B$
is also allowed, but it does not affect the following analysis in any way.
Taking into account the relation $\tr \mu_A^k =\tr \mu_B^k$ for $k \leq N-1$,
and that the equality in \eqref{eq:lowerbound} holds
when $|\mu_{A,1}|=\cdots=|\mu_{A,N}|$ and $|\mu_{B,1}|=\cdots=|\mu_{B,N}|$,
we conclude that the minimum of $V_{M}$ is realized at
\begin{equation}
\mu_A =\Lambda^2 \diag(1,\omega,\cdots, \omega^{N-1}),~~~~~\mu_B=0, \label{eq:vevmu}
\end{equation}
up to transformations by $\SU(N)_A$, where $\omega=e^{2 \pi i/N}$.
The value of $V_M$ at that point is
\begin{equation}
V_{M}|_{\rm min}=N|\lambda_A^2 \Lambda^4| . \label{eq:mimpot}
\end{equation}
%The equations \eqref{eq:VMeq2} and \eqref{eq:VMeq3} are now satisfied, and the equation \eqref{eq:VMeq1} can also be solved.

  The full potential becomes equal to $V_{M}$ if the $F$-term conditions of $\mu_{A,B}$ are satisfied.
That is, the inequality in \eqref{eq:Vinequality} is saturated if and only if
\bea
0&=&
%\frac{\partial W}{\partial \mu_A}=
-\lambda_A  M_A  +\sum_{k=2}^N {X_{k}} \left( \mu_A^{k-1} -\frac{\bf{1}}{N} \tr \mu_A^{k-1} \right), \label{FOO}\\
0&=&
%\frac{\partial W}{\partial \mu_B}=
\lambda_B  M_B  -\sum_{k=2}^N {X_{k}} \left( \mu_B^{k-1} -\frac{\bf{1}}{N} \tr \mu_B^{k-1} \right).\label{BAR}
\eea
From  \eqref{eq:vevmu} and \eqref{BAR}, we obtain $M_B=0$.
The equation \eqref{FOO} says that $M_A$ is diagonal and traceless.
%\begin{equation}
%M_A=\diag(M_{A,1}, \cdots, M_{A,N}),
%\end{equation} with $\sum_{i=1}^N M_{A,i}=0$.
$X_k$ can  then be solved, and are linear combinations of $M_{A,k}$.
%This is consistent thanks to $\sum_{i=1}^N M_{A,i}=\tr M_A=0$.

In summary, the pseudo-moduli space up to flavor rotations is parameterized by
the diagonal entries  of $M_A$,
\begin{equation}
M_A=\diag(M_{A,1}, \cdots, M_{A,N}),
\end{equation}
% $M_{A,k}$ for $k=1,\ldots, N$
with one constraint  $\sum_{i=1}^N M_{A,i}=0$, and other vevs can be expressed in terms of $M_{A,k}$.
The potential is
\begin{equation}
V=N|\lambda_A^2 \Lambda^4| \label{eq:mpot}
\end{equation}
 in this direction and in particular constant.

\subsection{Corrections to the pseudo-moduli}

So far we took
the K\"ahler potential of $M_{A,B}$ is canonical.
Let us take the %
quantum corrections to the K\"ahler potential into account.
We compute the corrections at the leading order of $\lambda_A \ll 1$.
We need to perform this computation in  the region where
$M_{A,i}$ are much larger than the dynamical scale of the $\SU(N)_C$ gauge theory,
to make sure that the potential shows no runaway behavior.
In this region, we can neglect the effect of $\SU(N)_C$ gauge interaction.

From the superpotential interaction \eqref{eq:treesuper}, we obtain the leading order correction to the effective action of $M_A$ as
\bea
-S_{\rm eff} \supset  |\lambda_A|^2 &\int& d^4x d^4 y (F_{M_A})^a_b(x) (F_{M_A}^*)^{\bar{a}}_{\bar{b}}(y)  \nonumber \\
&& \times \vev{ (\mu_A)_a^b(x)   (\mu^*_A)_{\bar{a}}^{\bar{b}}(y)}+\cdots, \label{eq:leadingcorrection}
\eea
where $F_{M_A}$ is the $F$-component of the chiral multiplet $M_A$, and the ellipsis denotes supersymmetric completion of the term explicitly written.
We are using Euclidean signature here.

To evaluate the expression above, we need to know the two-point function of two $\mu$'s.
When $ |x-y| \ll |M_A|^{-1}$,
the effect of the nonzero vev of $M_A$ can be neglected.
Then the correlation function is simply that of the $T_N$ theory.
Since $\mu_A$ has the scaling dimension two, we have
\begin{equation}
\vev{ (\mu_{A})_a^b(x) (\mu_{A}^*)_{\bar{a}}^{\bar{b}}(y)}\sim  c|x-y|^{-4} \delta^{a\bar{a}}\delta_{b\bar{b} }
\end{equation} with a positive coefficient $c$, thanks to the unitarity of the theory.
In fact $\mu_{A}$ sits in the conserved current supermultiplet, and the precise value of $c$ is known to be proportional to $N$ \cite{Gaiotto:2009we}.

When $|x-y|\gg |M_A|^{-1}$, $M_A$ acts as a mass deformation of the $T_N$ theory leading to a mass gap; this will lead to the behavior the two-point function:
\begin{equation}
\vev{ (\mu_{A})_a^b(x) (\mu_{A}^*)_{\bar{a}}^{\bar{b}}(y)}\sim \text{exponentially suppressed}\label{eq:correlation}.
\end{equation}
We will study this point in more detail below, and let us continue assuming its validity.

For the purpose of computing the potential, we may take $F_{M_A}$ to be independent of the space-time coordinates.
Then \eqref{eq:leadingcorrection} gives
\bea
-S_{\rm eff} &\supset  |\lambda_A|^2 \int_{\epsilon_{\rm UV} <|x-y|<|M_A|^{-1}} d^4x d^4 y  \tr (F^\dagger F) c |x-y|^{-4} \nonumber \\
&=-2\pi^2 c |\lambda_A|^2 \int d^4x  \tr (F^\dagger F)  \log (\epsilon_{\rm UV} |M_A|),
\eea
where $\epsilon_{\rm UV}$ is the UV cutoff.
Therefore, the kinetic term of $F_{M_A}$ including the tree level term and a counterterm
is given as $[1-2\pi^2 c |\lambda_A|^2 \log (\epsilon_{\rm UV} |M_A|)+\delta]\tr (F^\dagger F) $. Here $\delta$ is a counterterm of the kinetic term,
which is chosen to cancel the divergent term $\log \epsilon_{\rm UV}$.
Using this, the potential \eqref{eq:mpot} in the pseudo moduli space is modified to
\bea
V \simeq \left[1+2\pi^2 c |\lambda_A|^2 \log ( |M_A|/\mu_{\rm RG}) \right]N|\lambda_A^2 \Lambda^4|,\label{eq:logpotential}
\eea where $\mu_{\rm RG}$ is a renormalization scale determined by $\epsilon_{\rm UV}$ and $\delta$. This prefactor containing  large logarithm can be improved by renormalization group by taking $\mu_{\rm RG}$ to be of order $|M_A|$ and evaluating 
the parameters $\lambda_A$ and $\Lambda$ at this renormalization scale. See \cite{ArkaniHamed:1997ut} for details.
Reflection positivity of unitary quantum field theory
ensures that the constant $c$ is positive.
Then this potential is logarithmically increasing for large $|M_A|$, and hence there is no runaway behavior.

Finally let us discuss where the actual position of the vacuum is likely to be.
The most likely possibility is that the vacuum is at $M_A=0$, where the $\U(1)_R$ symmetry is unbroken.
Using the low energy effective field theory of $\mu_{A,B}$ and $M_{A,B}$, we can calculate the Coleman-Weinberg potential
near the point $M_A=0$, as was done in the $N=2$ case in \cite{Chacko:1998si}. We have not done this computation in our theory,
but the theorem of \cite{Shih:2007av} tells us that the potential minimum of this effective theory is at the point where the $\U(1)_R$ is not broken,
because $\mu_{A,B}$ has $R$-charge 0 and $M_{A,B}$ has $R$-charge 2. It is likely that this point is the global minimum
in the full theory.

%\appendix
\subsection{More on Eq.~\eqref{eq:correlation}}\label{detail}

Let us study the long distance behavior of the two-point function of $\mu$ \eqref{eq:correlation} in more detail.
Let us first consider the case of $N=2$, where $\mu_A \sim QQ$ is a composite of free chiral multiplets as mentioned above.
By the superpotential \eqref{eq:treesuper}, the $Q$ fields get masses of order $|M_A|$.
Then the correlation function $\vev{(QQ)(x) (QQ)^*(y)}$ is exponentially suppressed as $\exp(-|M_A||x-y|)$ at long distances.

For $N \geq 3$, we use an $\CN{=}2$ duality of the $T_N$ theory \cite{Gaiotto:2009we}.
The $\SU(N)_C$ gauge group plays no role in the discussion and we neglect it.
Let us consider the $T_N$ theory coupled to a quiver gauge group $\SU(N-1)-\SU(N-2)-\cdots-\U(1)$, where the $\SU(N-1)$ gauge fields
are coupled to the subgroup of one of the three $\SU(N)$'s, say $\SU(N)_C$. Then this theory is dual to the Lagrangian field theory
given by the quiver $\SU(N)_A -\SU(N)_1-\cdots - \SU(N)_{N-2}-\SU(N)_B$. The $\SU(N)_A$ and $\SU(N)_B$ are the flavor groups of the
$T_N$ theory, and $\SU(N)_i~(i=1,\cdots, N-2)$ are gauge groups. There are bifundamental hypermultiplets between two adjacent $\SU(N)$ gauge groups.
In the dual side, the operator $\mu_A$ is really given as a composite of quarks $Q, \tilde{Q}$ which are bifundamentals of $\SU(N)_A \times \SU(N)_1$.
So we have $\mu_A \sim Q\tilde{Q}$ and the vev of $M_A$ gives masses to these quarks. Then \eqref{eq:correlation}
is justified as in the $N=2$ case, as far as the coupling constants are all finite. %, although there is a subtlety which we will discuss later.
The weak coupling limit of $\SU(N-1)-\SU(N-2)-\cdots-\U(1)$ corresponds roughly to an infinite coupling limit of the dual theory,
but we expect no phase transition in the behavior in this limit because of the $\CN{=}2$ supersymmetry.

In the above calculation, we argued that the dual quarks $Q, \tilde{Q}$ get masses of order $|M_A|$ since they are coupled
as $\tr M_A Q \tilde{Q}$ in the superpotential. However, in the dual quiver theory, there is the adjoint chiral field $\Phi_1$ of the gauge group $\SU(N)_1$
which is coupled to the quarks as $\tr \Phi_1 \tilde{Q} Q$. If the vev of $\Phi_1$ happens to be such that some of the quarks are
almost massless, the correlation function at $|x-y| \gg |M_A|^{-1}$ behaves as $c' |x-y|^{-4}$. We consider it to be unlikely; note that in the deformed moduli space
\eqref{eq:deformedmoduli}, the Coulomb moduli fields of the $T_N$ theory are fixed somewhere, so the vev of $\Phi_1$ cannot be freely chosen.
Even if it happens, the constant $c'$ is smaller than $c$ because at least some of the quarks are massive and the number
of massless quarks are reduced in the infrared (IR). Then one can see that there is a UV contribution to the kinetic term of $F_{M_A}$
proportional to $(c-c')\log(|M_A|)$ with $c-c'>0$. Assuming that the IR contribution is not so bad as to spoil this UV contribution,
the logarithmic increasing of the potential is not changed. For example, if the correlation function is cutoff in the IR at the dynamical scale $\Lambda$,
then \eqref{eq:logpotential} is valid just by replacing $c \to c-c'$.

%%%%%%%%%%%%%%%%%%%%%%%%%%%%%%%%%%%%  acknowledgements
\section*{Acknowledgments}
The authors would like to thank Professor Y. Kikukawa for suggesting us to consider this problem, when Y.~T.~gave a talk on the authors' previous works in University of Tokyo, Komaba.
K.~M.~and W.~Y.~would like to thank Simons Summer Workshop 2013 in Mathematics and Physics, 
where this work was finished, for hospitality.
Y.~T.~also thanks
the Aspen Center for Physics and the NSF Grant \#1066293 for hospitality while the manuscript was finalized. 
The work of K.M. is supported by a JSPS postdoctoral fellowship for research abroad. The work of  Y.T. is supported in part by World Premier International Research Center Initiative  (WPI Initiative),  MEXT, Japan through the Institute for the Physics and Mathematics of the Universe, the University of Tokyo. The work of W.Y. is supported in part by the Sherman Fairchild scholarship and by DOE grant DE-FG02-92-ER40701.
The work of K.Y. is supported in part by NSF grant PHY-0969448.

%\appendix
%%%%%%%%%%%%%%%%%%%%%%%%%%%%%%%%%%%%%%%%%%%%%%%%%%%%%
%\section{}
%\label{appsec:}

%\bibliographystyle{ytphys}
%\small\baselineskip=.9\baselineskip
\bibliography{ref}

\begin{thebibliography}{8}%
\makeatletter
\providecommand \@ifxundefined [1]{%
 \@ifx{#1\undefined}
}%
\providecommand \@ifnum [1]{%
 \ifnum #1\expandafter \@firstoftwo
 \else \expandafter \@secondoftwo
 \fi
}%
\providecommand \@ifx [1]{%
 \ifx #1\expandafter \@firstoftwo
 \else \expandafter \@secondoftwo
 \fi
}%
\providecommand \natexlab [1]{#1}%
\providecommand \enquote  [1]{``#1''}%
\providecommand \bibnamefont  [1]{#1}%
\providecommand \bibfnamefont [1]{#1}%
\providecommand \citenamefont [1]{#1}%
\providecommand \href@noop [0]{\@secondoftwo}%
\providecommand \href [0]{\begingroup \@sanitize@url \@href}%
\providecommand \@href[1]{\@@startlink{#1}\@@href}%
\providecommand \@@href[1]{\endgroup#1\@@endlink}%
\providecommand \@sanitize@url [0]{\catcode `\\12\catcode `\$12\catcode
  `\&12\catcode `\#12\catcode `\^12\catcode `\_12\catcode `\%12\relax}%
\providecommand \@@startlink[1]{}%
\providecommand \@@endlink[0]{}%
\providecommand \url  [0]{\begingroup\@sanitize@url \@url }%
\providecommand \@url [1]{\endgroup\@href {#1}{\urlprefix }}%
\providecommand \urlprefix  [0]{URL }%
\providecommand \Eprint [0]{\href }%
\providecommand \doibase [0]{http://dx.doi.org/}%
\providecommand \selectlanguage [0]{\@gobble}%
\providecommand \bibinfo  [0]{\@secondoftwo}%
\providecommand \bibfield  [0]{\@secondoftwo}%
\providecommand \translation [1]{[#1]}%
\providecommand \BibitemOpen [0]{}%
\providecommand \bibitemStop [0]{}%
\providecommand \bibitemNoStop [0]{.\EOS\space}%
\providecommand \EOS [0]{\spacefactor3000\relax}%
\providecommand \BibitemShut  [1]{\csname bibitem#1\endcsname}%
\let\auto@bib@innerbib\@empty
%</preamble>
\bibitem [{\citenamefont {Gaiotto}(2012)}]{Gaiotto:2009we}%
  \BibitemOpen
  \bibfield  {author} {\bibinfo {author} {\bibfnamefont {D.}~\bibnamefont
  {Gaiotto}},\ }\href {\doibase 10.1007/JHEP08(2012)034} {\bibfield  {journal}
  {\bibinfo  {journal} {JHEP}\ }\textbf {\bibinfo {volume} {1208}},\ \bibinfo
  {pages} {034} (\bibinfo {year} {2012})},\ \Eprint
  {http://arxiv.org/abs/0904.2715} {arXiv:0904.2715 [hep-th]} \BibitemShut
  {NoStop}%
%%CITATION = ARXIV:0904.2715;%%
\bibitem [{\citenamefont {Izawa}\ and\ \citenamefont
  {Yanagida}(1996)}]{Izawa:1996pk}%
  \BibitemOpen
  \bibfield  {author} {\bibinfo {author} {\bibfnamefont {K.-I.}\ \bibnamefont
  {Izawa}}\ and\ \bibinfo {author} {\bibfnamefont {T.}~\bibnamefont
  {Yanagida}},\ }\href {\doibase 10.1143/PTP.95.829} {\bibfield  {journal}
  {\bibinfo  {journal} {Prog.Theor.Phys.}\ }\textbf {\bibinfo {volume} {95}},\
  \bibinfo {pages} {829} (\bibinfo {year} {1996})},\ \Eprint
  {http://arxiv.org/abs/hep-th/9602180} {arXiv:hep-th/9602180 [hep-th]}
  \BibitemShut {NoStop}%
%%CITATION = HEP-TH/9602180;%%
\bibitem [{\citenamefont {Intriligator}\ and\ \citenamefont
  {Thomas}(1996)}]{Intriligator:1996pu}%
  \BibitemOpen
  \bibfield  {author} {\bibinfo {author} {\bibfnamefont {K.~A.}\ \bibnamefont
  {Intriligator}}\ and\ \bibinfo {author} {\bibfnamefont {S.~D.}\ \bibnamefont
  {Thomas}},\ }\href {\doibase 10.1016/0550-3213(96)00261-1} {\bibfield
  {journal} {\bibinfo  {journal} {Nucl.Phys.}\ }\textbf {\bibinfo {volume}
  {B473}},\ \bibinfo {pages} {121} (\bibinfo {year} {1996})},\ \Eprint
  {http://arxiv.org/abs/hep-th/9603158} {arXiv:hep-th/9603158 [hep-th]}
  \BibitemShut {NoStop}%
%%CITATION = HEP-TH/9603158;%%
\bibitem [{\citenamefont {Maruyoshi}\ \emph {et~al.}(2013)\citenamefont
  {Maruyoshi}, \citenamefont {Tachikawa}, \citenamefont {Yan},\ and\
  \citenamefont {Yonekura}}]{Maruyoshi:2013hja}%
  \BibitemOpen
  \bibfield  {author} {\bibinfo {author} {\bibfnamefont {K.}~\bibnamefont
  {Maruyoshi}}, \bibinfo {author} {\bibfnamefont {Y.}~\bibnamefont
  {Tachikawa}}, \bibinfo {author} {\bibfnamefont {W.}~\bibnamefont {Yan}}, \
  and\ \bibinfo {author} {\bibfnamefont {K.}~\bibnamefont {Yonekura}},\
  }\href@noop {} {\  (\bibinfo {year} {2013})},\ \Eprint
  {http://arxiv.org/abs/1305.5250} {arXiv:1305.5250 [hep-th]} \BibitemShut
  {NoStop}%
%%CITATION = ARXIV:1305.5250;%%
\bibitem [{\citenamefont {Seiberg}(1994)}]{Seiberg:1994bz}%
  \BibitemOpen
  \bibfield  {author} {\bibinfo {author} {\bibfnamefont {N.}~\bibnamefont
  {Seiberg}},\ }\href {\doibase 10.1103/PhysRevD.49.6857} {\bibfield  {journal}
  {\bibinfo  {journal} {Phys.Rev.}\ }\textbf {\bibinfo {volume} {D49}},\
  \bibinfo {pages} {6857} (\bibinfo {year} {1994})},\ \Eprint
  {http://arxiv.org/abs/hep-th/9402044} {arXiv:hep-th/9402044 [hep-th]}
  \BibitemShut {NoStop}%
%%CITATION = HEP-TH/9402044;%%
\bibitem [{\citenamefont {Arkani-Hamed}\ and\ \citenamefont
  {Murayama}(1998)}]{ArkaniHamed:1997ut}%
  \BibitemOpen
  \bibfield  {author} {\bibinfo {author} {\bibfnamefont {N.}~\bibnamefont
  {Arkani-Hamed}}\ and\ \bibinfo {author} {\bibfnamefont {H.}~\bibnamefont
  {Murayama}},\ }\href {\doibase 10.1103/PhysRevD.57.6638} {\bibfield
  {journal} {\bibinfo  {journal} {Phys.Rev.}\ }\textbf {\bibinfo {volume}
  {D57}},\ \bibinfo {pages} {6638} (\bibinfo {year} {1998})},\ \Eprint
  {http://arxiv.org/abs/hep-th/9705189} {arXiv:hep-th/9705189 [hep-th]}
  \BibitemShut {NoStop}%
%%CITATION = HEP-TH/9705189;%%
\bibitem [{\citenamefont {Chacko}\ \emph {et~al.}(1998)\citenamefont {Chacko},
  \citenamefont {Luty},\ and\ \citenamefont {Ponton}}]{Chacko:1998si}%
  \BibitemOpen
  \bibfield  {author} {\bibinfo {author} {\bibfnamefont {Z.}~\bibnamefont
  {Chacko}}, \bibinfo {author} {\bibfnamefont {M.~A.}\ \bibnamefont {Luty}}, \
  and\ \bibinfo {author} {\bibfnamefont {E.}~\bibnamefont {Ponton}},\
  }\href@noop {} {\bibfield  {journal} {\bibinfo  {journal} {JHEP}\ }\textbf
  {\bibinfo {volume} {9812}},\ \bibinfo {pages} {016} (\bibinfo {year}
  {1998})},\ \Eprint {http://arxiv.org/abs/hep-th/9810253}
  {arXiv:hep-th/9810253 [hep-th]} \BibitemShut {NoStop}%
%%CITATION = HEP-TH/9810253;%%
\bibitem [{\citenamefont {Shih}(2008)}]{Shih:2007av}%
  \BibitemOpen
  \bibfield  {author} {\bibinfo {author} {\bibfnamefont {D.}~\bibnamefont
  {Shih}},\ }\href {\doibase 10.1088/1126-6708/2008/02/091} {\bibfield
  {journal} {\bibinfo  {journal} {JHEP}\ }\textbf {\bibinfo {volume} {0802}},\
  \bibinfo {pages} {091} (\bibinfo {year} {2008})},\ \Eprint
  {http://arxiv.org/abs/hep-th/0703196} {arXiv:hep-th/0703196 [hep-th]}
  \BibitemShut {NoStop}%
%%CITATION = HEP-TH/0703196;%%
\end{thebibliography}%

\end{document}